\documentclass[preprints,article,accept,moreauthors,pdftex,10pt,a4paper]{mdpi}

\firstpage{1} 
\pubvolume{xx}
\issuenum{1}
\articlenumber{5}
\pubyear{2018}
\copyrightyear{2018}
\history{Submitted: November 6, 2018}

\Title{HIJING++, a Heavy Ion Jet INteraction Generator for the High-luminosity Era of the LHC and Beyond$^{\dagger}$}

\Author{Gábor Bíró$^{1,2}$\orcidA{}, Gábor Papp$^{1}$\orcidB{}, Gergely Gábor Barnaföldi$^{2}$\orcidC{}, Dániel Nagy$^{1}$,
        Miklos Gyulassy$^{2,3,4,5}$\orcidD{}, P\'eter L\'evai$^{2}$, Xin-Nian Wang$^{3,4}$\orcidE{} and Ben-Wei Zhang$^{3}$\orcidF{}
}

\AuthorNames{Gábor Bíró, Gábor Papp, Gergely Gábor Barnaföldi, Dániel Nagy, Miklos Gyulassy, Péter Lévai, Xin-Nian Wang and Ben-Wei Zhang}

\address{%
$^{1}$ \quad Institute for Physics, E\"otv\"os Lor\'and University, 1/A P\'azm\'any P. S\'et\'any, H-1117, Budapest, Hungary\\
$^{2}$ \quad Wigner Research Centre for Physics of the Hungarian Academy of Sciences, 29-33 Konkoly-Thege Mikl\'os Str, H-1121 Budapest, Hungary\\
$^{3}$ \quad Key Laboratory of Quark \& Lepton Physics (MOE) and Institute of Particle Physics, Central China Normal University, Wuhan 430079, China\\
$^{4}$ \quad Nuclear Science Division, MS 70R0319, Lawrence Berkeley National Laboratory, Berkeley, California 94720 USA \\
$^{5}$ \quad Pupin Lab MS-5202, Department of Physics, Columbia University, New York, NY 10027, USA
}

\corres{Correspondence: biro.gabor@wigner.mta.hu; Tel.: +36-30-308-6742}

\firstnote{Presented at Hot Quarks 2018 - Workshop for young scientists on the physics of ultrarelativistic nucleus-nucleus collisions, Texel, The Netherlands, September 7-14 2018}

\abstract{
\texttt{HIJING++} (Heavy Ion Jet INteraction Generator) is the successor of the widely used original \texttt{HIJING}~\cite{HIJING, HIJING2}, developed almost three decades ago. While the old versions (1.x and 2.x) were written in \texttt{FORTRAN}, \texttt{HIJING++} was completely rewritten in C++. During the development we keep in mind the requirements of the high-energy heavy-ion community: the new Monte Carlo software have a well designed modular framework, therefore any future modifications are much easier to implement. It contains all the physical models that were also present in it’s predecessor, but utilizing modern C++ features it also includes native thread based parallelism, an easy-to-use analysis interface and a modular plugin system, which makes room for possible future improvements. In this paper we summarize the results of our performance tests measured on 2 widely used architectures.
}



\begin{document}

\section{Introduction}

During the approaching Long Shutdown 2 (LS2) of the Large Hadron Collider (LHC) in 2019-2020 many technical improvement will occur in the accelerator complex, in the detector and in the data acquisition systems. These will result in a huge increase of the number of expected collisions per second and also the amount of measured data per event will grow rapidly. This period is the forerunner of the next generation of particle accelerators, such as the High-Luminosity LHC (HL-LHC) or the Future Circular Collider (FCC), where we will accumulate high-energy experimental data in a higher rate than ever. In parallel we need to improve also the numerical tools in order to be able to keep up the requisites of the high-precision era.

The new \texttt{HIJING++} heavy-ion Monte Carlo framework is written from scratch with a modular, effective C++ structure and with built-in CPU based parallelism in order to fulfill these requirements. Though the program flow is based on the original \texttt{FORTRAN HIJING}\cite{HIJING, HIJING2}, the design is completely revised so the main components of the program can work together effectively. Such components are the most recent versions of \texttt{PYTHIA8}~\cite{PYTHIA} (used for the hard scattering processes and for the hadronization), \texttt{LHAPDF6}~\cite{LHAPDF}, the \texttt{GNU Scientific Library}~\cite{VEGAS, GSL} (utilizing the VEGAS multi-dimensional Monte Carlo integration), and the \texttt{CERN ROOT}~\cite{ROOT} data analysis software along with the \texttt{HijAnalysis} data collector framework. 

\texttt{HIJING++} is intended to work effectively regarding different aspects, not just based on the raw performance of the CPU. As an example, it is possible to replace any of the main components, such as the jet quenching and shadowing algorithms, in a convenient, well defined way, without modifying the core code. An another built-in feature is the above mentioned \texttt{HijAnalysis} framework, which adds the possibility to define any kind of data collecting objects, such as \texttt{ROOT TTree}s, histograms or simple ASCII files to collect all final state particles event-by-event. Utilizing modern C++ features, the result of a run will be data structures that can be further processed in a convenient way.

In the following section we present the results of the performance tests of the pre-release version of \texttt{HIJING++}, taking advantage of these features.

\section{Results}

We have already presented preliminary physics and performance results in Ref \cite{HPP1,HPP2}. Here we summarize the benchmark tests measured on two different machines.

\subsection{Benchmark setups}

In order to measure the performance in a real case situation, we calculated 6 different histograms to collect various quantities of the current run, such as the impact parameter, number of binary collisions, event multiplicity, $p_T$ spectra and pseudorapidity distributions of different identified hadrons with various binnings. We performed each run several times in order to reduce fluctuations. The main parameters of the different run setups are summarized in Table \ref{tab:setups}~\cite{PDF1, PDF2}.

\begin{table}[H]
\caption{The run setups and their main parameters.}
\label{tab:setups}
\centering
\begin{tabular}{ccc}
\toprule
\textbf{Collision system}	& \textbf{Event number}	& \textbf{(n)PDF}\\
\midrule
pp		    & $10^6$	& \texttt{CT14nlo}~\cite{PDF2}\\
p-Pb		& $10^4$	& \texttt{CT14nlo} (for protons), \texttt{EPPS16nlo\_CT14nlo\_Pb208}~\cite{PDF1} (for lead nuclei) \\
Pb-Pb		& $10^3$	& \texttt{EPPS16nlo\_CT14nlo\_Pb208} \\
\bottomrule
\end{tabular}
\end{table}

The tests were made on 2 commonly used, typical architectures, whose parameters are listed in Table \ref{tab:architectures}~\cite{INTEL}. These setups represents common use cases in the heavy-ion community: CPUs with lower TDP values (\textit{thermal design power} - the higher the value, the larger the power consumption and performance) and its variants are widely used in recent laptops and ultrabooks, while CPUs with higher TDP are common in desktop computers or larger workstations, clusters.

\begin{table}[H]
\caption{The computer architectures~\cite{INTEL} used to measure run performance.}
\label{tab:architectures}
\centering
\tablesize{\footnotesize}
\begin{tabular}{cccccc}
\toprule
\textbf{CPU type}	& \textbf{Release year}	& \textbf{Number of cores (threads)} & \textbf{Base (turbo) frequency} & \textbf{TDP} & \textbf{RAM}\\
\midrule
Intel(R) Core(TM) i5-8250U  & Q3'17 & 4 (8) & 1.6 GHz (3.4 GHz) & 15 W & 8 GB \\
Intel(R) Xeon(TM) E3-1231 v3 & Q2'14 & 4 (8) & 3.4 GHz (3.8 GHz) & 80 W & 32 GB \\
\bottomrule
\end{tabular}
\end{table}

\subsection{Results}

\begin{figure}[H]
\centering
\includegraphics[width=0.325\textwidth]{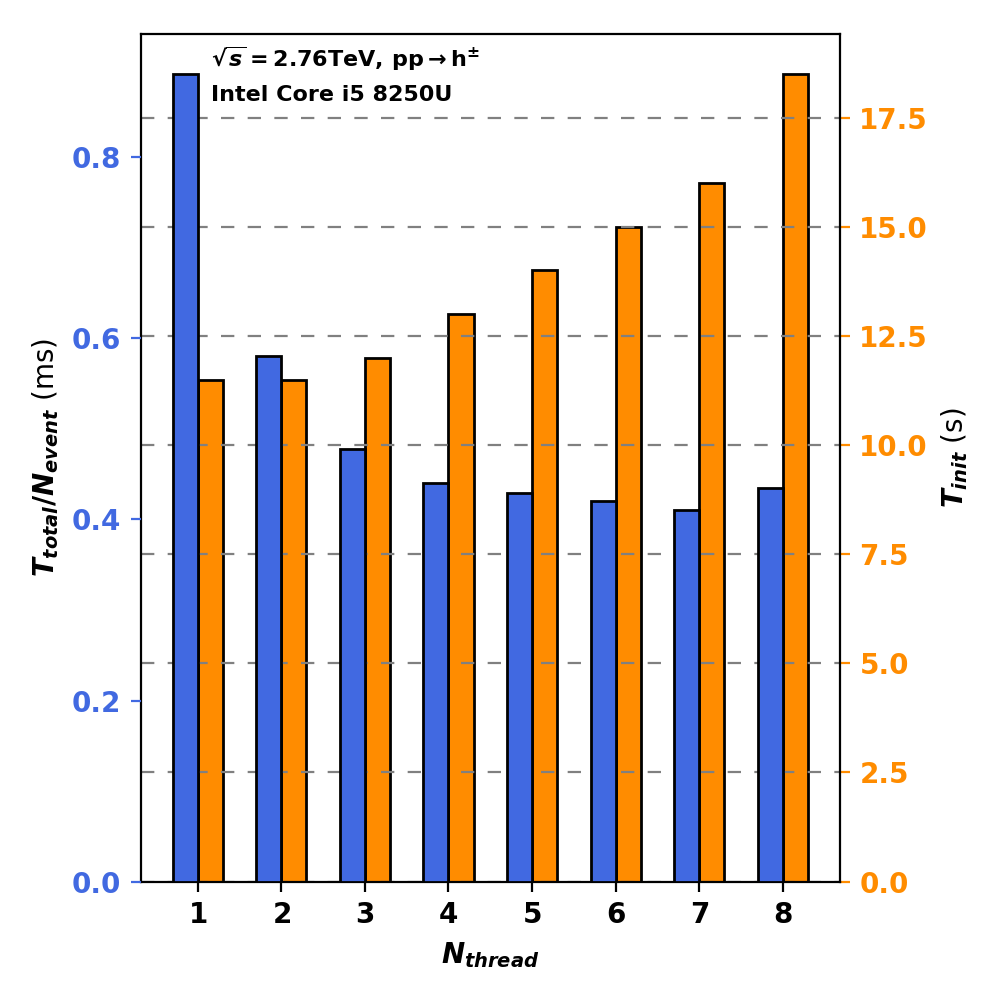}
\includegraphics[width=0.325\textwidth]{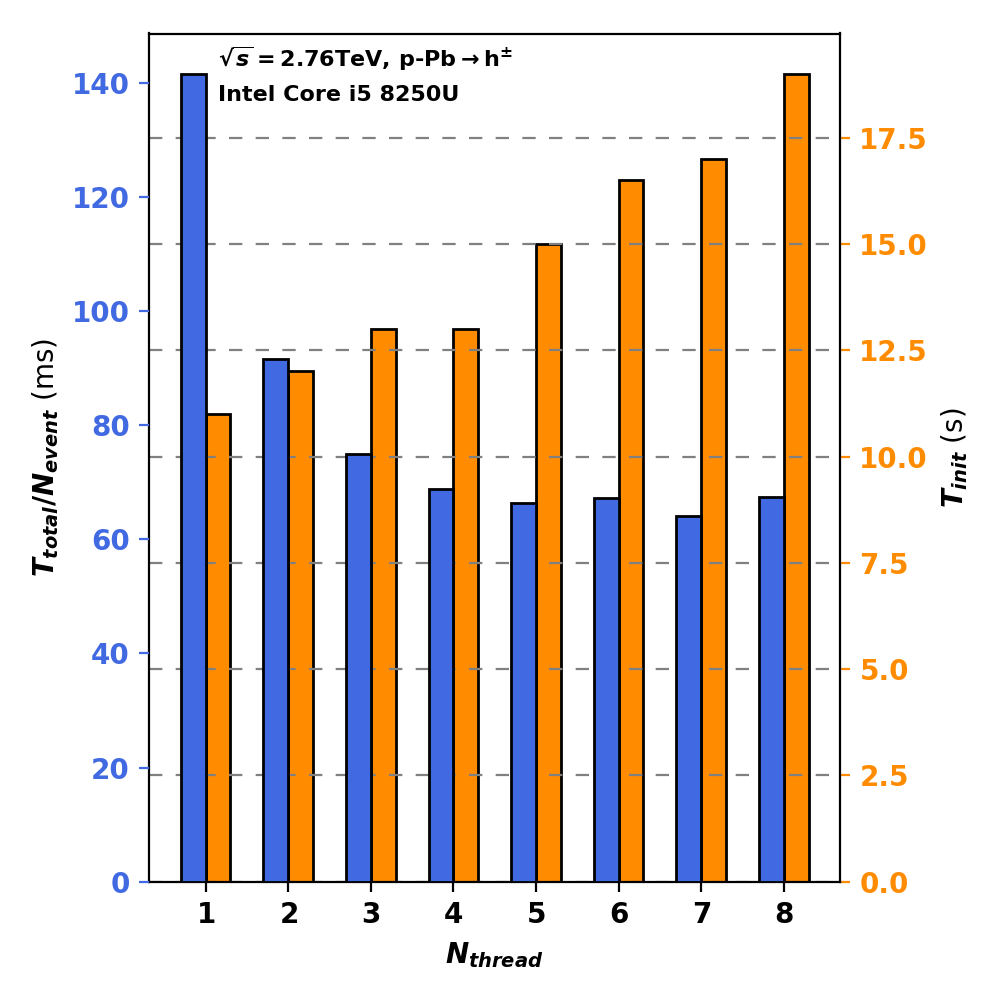}
\includegraphics[width=0.325\textwidth]{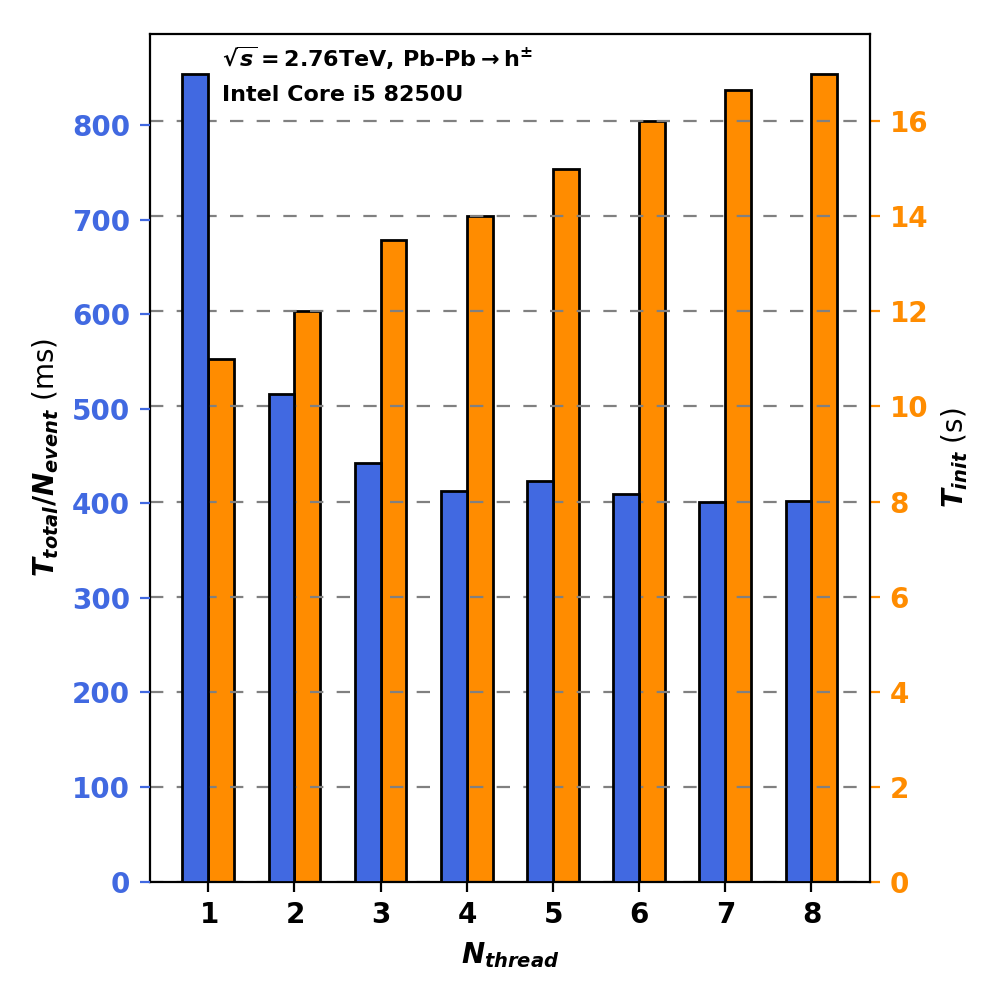}
\includegraphics[width=0.325\textwidth]{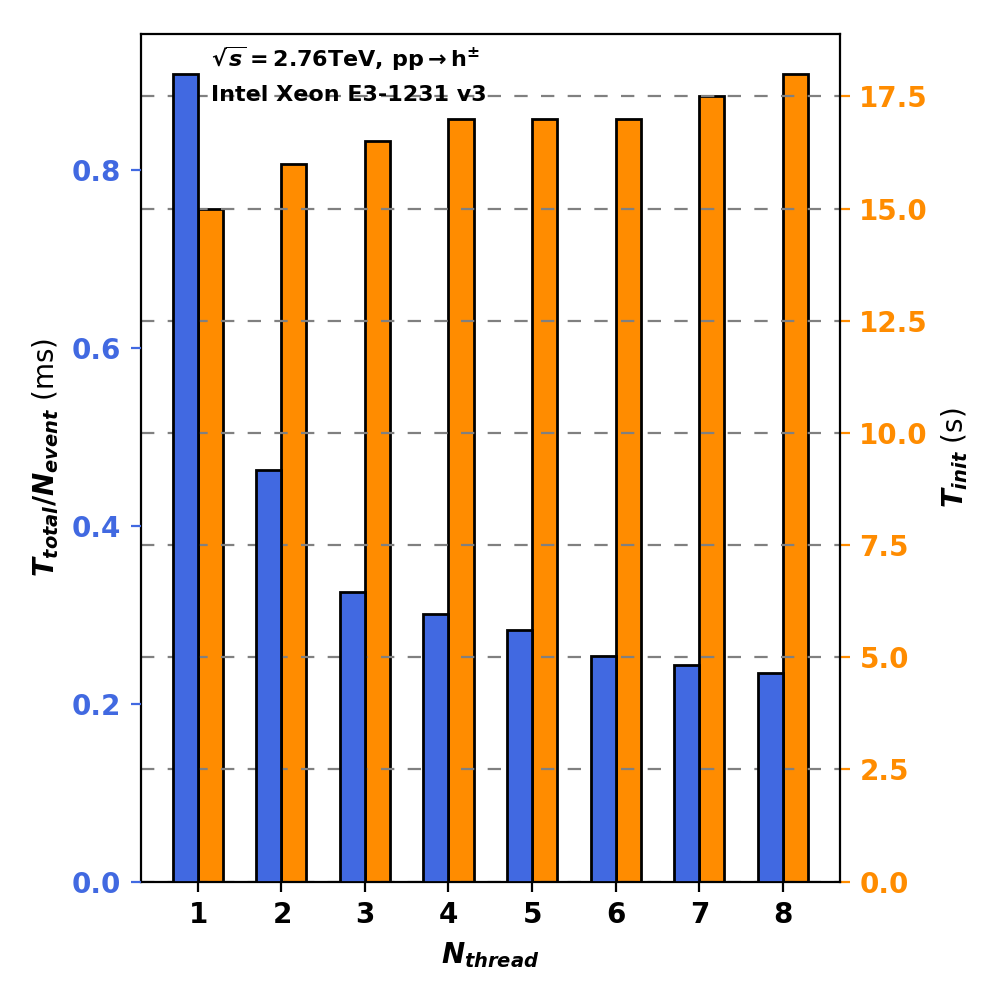}
\includegraphics[width=0.325\textwidth]{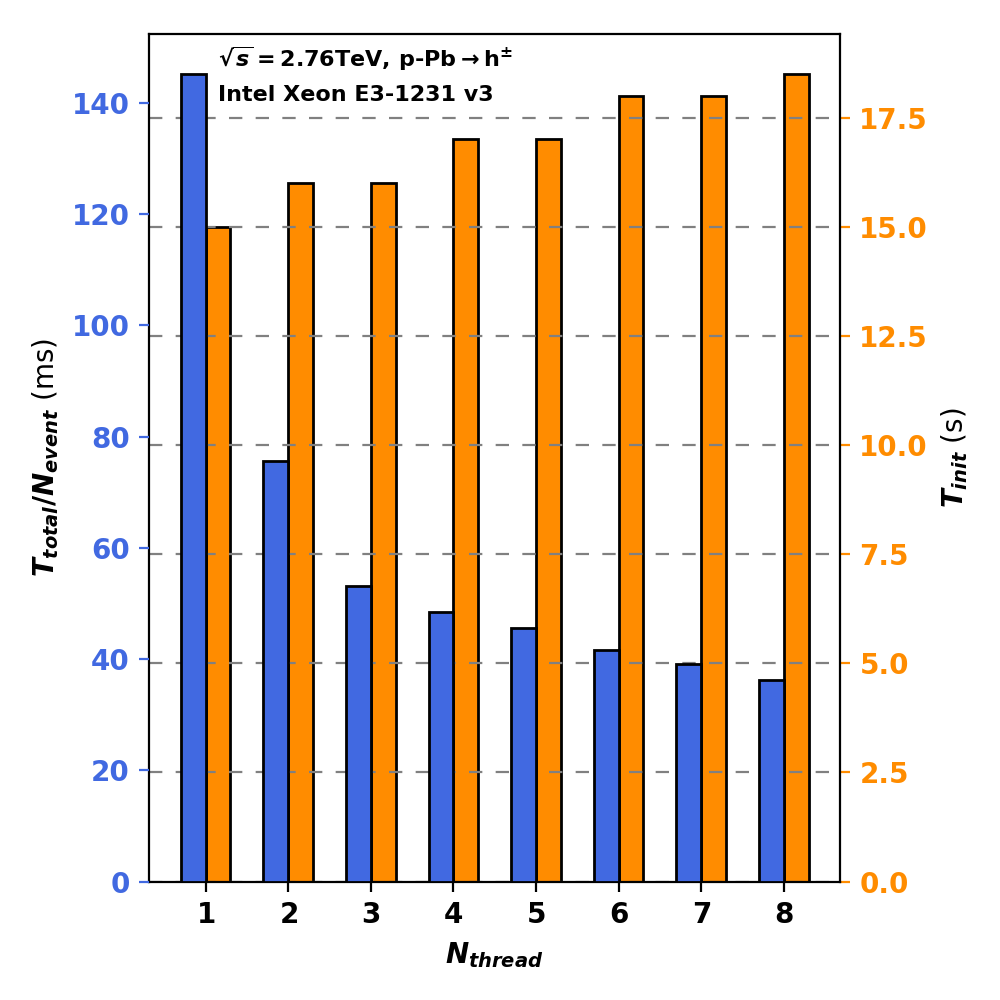}
\includegraphics[width=0.325\textwidth]{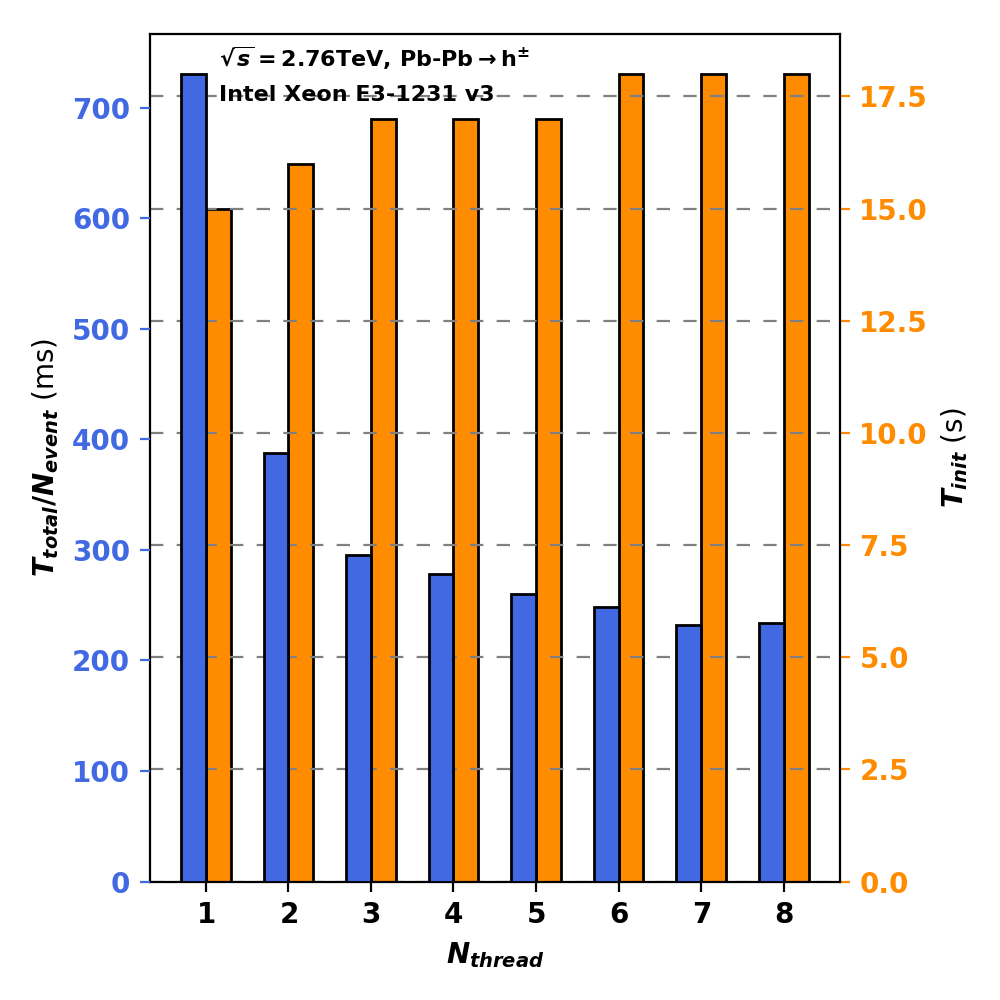}
\caption{The total runtime normalized with the event number (blue bars) and the initialization time (orange bars) versus the CPU threads, measured on an Intel(R) Core(TM) i5-8250U CPU (\textbf{upper panels}) and on an Intel(R) Xeon(TM) E3-1231 v3 CPU (\textbf{lower panels}). The run parameters: $\sqrt{s}=2.76$ ATeV proton-proton (\textbf{left panels}), proton-lead (\textbf{middle panels}) and lead-lead collisions (\textbf{right panels}), using (nuclear) parton distribution functions \texttt{CT14nlo} (for protons) and \texttt{EPPS16nlo\_CT14nlo\_Pb208} (for lead nuclei), defining 6 different histogram analysis objects.}
\label{fig:cpu1-2}
\end{figure}

The results of the benchmarking runs for the two different CPUs are shown on Figure \ref{fig:cpu1-2}. As expected, the measured times show significant differences between the two system: using the CPU with the lower TDP value (\textit{upper panels}) by increasing the number of threads the total runtime decreases significally until $N_{thread}=4$, then the speedup gained from the multiple threads is compensated by the fact that more CPU cores have to share the same amount of energy, resulting in a decrease of the CPU frequency. In accorddance with this, the initialization time increases slightly with the increasing thread number. In contrast to these, on the \textit{lower panels} the results achieved with the higher performance desktop/server CPU are shown, where the speedup is more significant with the higher number of threads. In this case, the initialization time increases with a much lower rate. The reason is that this CPU doesn't have to decrease the performance when we are operating with multiple cores.

By fitting the measured results with Amdahl's law~\cite{AMDAHL} we can determine the maximum theoretical speedup compared to the single thread run that can be achieved on the specific architecture:
\begin{equation}
\textrm{Speedup}(N_{threads})=\frac{N_{threads}}{1+\alpha(N_{threads}-1)} \ \ \ ,
\end{equation}
where $\alpha$ is the non-parallelizable part of the code. According to the results summarized in Table \ref{tab:amdahl} the scalability on the higher performance CPU is better, the non-parallelizable parts (such as the thread managing system itself) result in a lower $\alpha$ value. However, using 3-4 threads \texttt{HIJING++} runs more efficiently also with the low TDP CPU, resulting in a considerably reduced runtime.

\begin{table}[H]
\caption{The maximum theoretical speedup compared to the single thread and the non-parallelizable part $\alpha$ of the code.}
\label{tab:amdahl}
\centering
\begin{tabular}{c|ccc|ccc}
\toprule
\multirow{2}{*}{\textbf{CPU}}	& \multicolumn{3}{c|}{\textbf{Speedup}} & \multicolumn{3}{c}{\textbf{$\alpha$}} \\
                                & pp & p-Pb & Pb-Pb & pp & p-Pb & Pb-Pb \\
\midrule
Intel(R) Core(TM) i5-8250U   & 2.6 & 2.7 & 2.6 & 0.38 & 0.37 & 0.38 \\
Intel(R) Xeon(TM) E3-1231 v3 & 6.4 & 6.6 & 4.5 & 0.16 & 0.15 & 0.22 \\
\bottomrule
\end{tabular}
\end{table}

In order put the performance of \texttt{HIJING++} into context, we measured and compared the (single thread) runtime of \texttt{PYTHIA8.2} and \texttt{HIJING v2.552}. We found that \texttt{HIJING++} is $\sim 30\%$ faster than \texttt{PYTHIA8.2} and $\sim 50\%$ slower than \texttt{HIJING v2.552}. 

This is not a surprising result, because the published \texttt{FORTRAN HIJING} was originally written with single precision floating point numbers: on one hand, this can lead to significant numerical errors (especially at LHC energies) when performing calculations with frequently occurring small quantities like $\sim \frac{m_q}{\sqrt{s}}\ll 1$, where $m_q$ is the mass of a given quark species and $\sqrt{s}$ is the center-of-mass energy. On the other hand, we measured the effect of modifying the \texttt{FORTRAN HIJING} into double precision, and we found that in such case it's runtime scales up by a factor of 4.

\section{Summary and conclusions}

We presented the results of the performance benchmarks of the new \texttt{HIJING++} heavy-ion Monte Carlo event generator using different CPUs and collision systems. Utilizing the built-in CPU parallelization and analysis frameworks \texttt{HIJING++} provides a significant decrease in the necessary computation time which is especially important at higher performance architectures. In the future developments further optimizations are planned to improve the scalability.

\vspace{6pt} 



\authorcontributions{G.B. developed the software framework and wrote the first version of the manuscript. Authors G.P., G.G.B., M.G., X.N.W., B.W.Z. and P.L. supervised the development, provided theoretical background and reviewed the manuscript. D.N. developed the benchmarking framework.}

\funding{This research was funded by Hungarian-Chinese cooperation grant No. MOST 2014DFG02050 and Wigner HAS-OBOR-CCNU grant; OTKA grants K120660, K123815, THOR COST action CA15213. Author G.B. acknowledge the support of Wigner Data Center and Wigner GPU Laboratory.}


\conflictsofinterest{The authors declare no conflict of interest.} 

\reftitle{References}



\begin{thebibliography}{999}

\bibitem{HIJING} X.N. Wang, M. Gyulassy,  {\em Phys. Rev.} {\bf 1991}, {\em D44}, 3501

\bibitem{HIJING2} W.T. Deng, X.N. Wang, R. Xu, {\em Phys. Rev.} {\bf 2011}, {\em C83}, 014915

\bibitem{PYTHIA} T. Sj\"ostrand, {\em Comput. Phys. Commun.} {\bf 2015},  {\em 191}, 159
  
\bibitem{LHAPDF} A. Buckley, J. Ferrando, S. Lloyd, K. Nordstr\"om, B. Page, M. R\"ufenacht, M. Sch\"onherr, G. Watt,  {\em Eur. Phys. J.} {\bf 2015}, {\em C75} 3, 132

\bibitem{GSL} M. Galassi et al, {\em GNU Scientific Library Reference Manual} (3rd Ed.), ISBN 0954612078

\bibitem{VEGAS} G.P. Lepage, {\em J. of Comp. Phys.} {\bf 1978}, {\em 27}, 192–203

\bibitem{ROOT} https://root.cern.ch/ (25.10.2018.)

\bibitem{HPP1} G.G. Barnaföldi, G. Bíró, M. Gyulassy, S.M. Haranozó, P. Lévai, G. Ma, G. Papp, X.N.Wang, B.W. Zhang, {\em Nucl. and Part. Phys. Proc.}  {\bf 2017}, {\em 289–290}, 373-376.

\bibitem{HPP2} G. Papp, G.G. Barnaföldi, G. Bíró, M. Gyulassy, Sz.M.Harangozó, G. Ma, P. Lévai, X.N. Wang, B.W. Zhang, , {\em Accepted to P.o.S.} {\bf 2018}, arXiv:1805.02635

\bibitem{PDF2} S. Dulat, T.J. Hou, J. Gao, M. Guzzi, J. Huston, P. Nadolsky, J. Pumplin, C. Schmidt, D. Stump, C. P. Yuan, {\em Phys. Rev. D} {\bf 2016}, {\em 93}, 033006

\bibitem{PDF1} K.J. Eskola, P. Paakkinen, H. Paukkunen, C.A. Salgado, {\em Eur. Phys. J.} {\bf 2017}, {\em C77} 3, 163

\bibitem{INTEL} https://ark.intel.com/products/80910/Intel-Xeon-Processor-E3-1231-v3-8M-Cache-3-40-GHz- (25.10.2018.), https://ark.intel.com/products/124967/Intel-Core-i5-8250U-Processor-6M-Cache-up-to-3-40-GHz- (25.10.2018.)

\bibitem{AMDAHL} G.M. Amdahl, {\em AFIPS Conference Proceedings} {\em 30}, 483.

\end{thebibliography}



\end{document}